\documentclass[14pt,reprint,twocolumn,longbibliography,
superscriptaddress,
amsmath,
amssymb,
aps,
pra
]{revtex4-1}
\usepackage{graphicx,amsmath,subfigure}%
\usepackage{color,xcolor}
\usepackage[bookmarks=false]{hyperref}
\hypersetup{colorlinks=true,citecolor=cyan,linkcolor=blue,urlcolor=blue,pdfstartview=FitH,bookmarksopen=true}
\usepackage{epstopdf}

\begin{document}

\title{Quantum computation in silicon-vacancy centers based on nonadiabatic geometric gates protected by dynamical decoupling}

\author{M.-R. Yun}
\affiliation{ School of Physics and Microelectronics, Key Laboratory of Materials Physics of Ministry of Education,
	Zhengzhou University, Zhengzhou 450052, China}
 \author{Jin-Lei Wu}
\affiliation{ School of Physics and Microelectronics, Key Laboratory of Materials Physics of Ministry of Education,
	Zhengzhou University, Zhengzhou 450052, China}
\author{L.-L. Yan}
\affiliation{ School of Physics and Microelectronics, Key Laboratory of Materials Physics of Ministry of Education,
	Zhengzhou University, Zhengzhou 450052, China}
\author{Yu Jia}
\email{jiayu@zzu.edu.cn}
 \affiliation{ Key Laboratory for Special Functional Materials of Ministry of Education, and School of Materials and Engineering, Henan University, Kaifeng 475001, China}	
 \affiliation{ School of Physics and Microelectronics, Key Laboratory of Materials Physics of Ministry of Education,
	Zhengzhou University, Zhengzhou 450052, China} 
 \affiliation{Institute of Quantum Materials and Physics, Henan Academy of Sciences, Zhengzhou 450046, China}
\author{S.-L. Su}
\email{slsu@zzu.edu.cn}
\author{C.-X. Shan}
\email{cxshan@zzu.edu.cn}
\affiliation{ School of Physics and Microelectronics, Key Laboratory of Materials Physics of Ministry of Education,
	Zhengzhou University, Zhengzhou 450052, China}

\begin{abstract}

Due to strong zero-phonon line emission, narrow inhomogeneous broadening, and stable optical transition frequencies, the quantum system consisting of negatively charged silicon-vacancy~(SiV) centers in diamond is highly expected to develop universal quantum computation. 
We propose to implement quantum computation for the first time using SiV centers placed in a one-dimensional phononic waveguide, for which quantum gates are realized in a nonadiabatic geometric way and protected by dynamical decoupling~(DD). The scheme has the feature of geometric quantum computation that is robust to control errors and the advantage of DD that is insensitive to environmental impact. Furthermore, the encoding of qubits in long-lifetime ground states of silicon-vacancy centers can reduce the effect of spontaneous emission. 
Numerical simulations demonstrate the practicability of the SiV center system for quantum computation and the robustness improvement of quantum gates by DD pulses. This scheme may provide a promising path toward high-fidelity geometric quantum computation in solid-state systems.

\end{abstract}

\maketitle

\section{Introduction}

Diamond color centers have great potential in various applications, such as quantum computation~\cite{doi:10.1063/5.0007444,PhysRevA.105.012611,PhysRevA.100.052332,Zhou:15,PhysRevA.85.042306}, state detection~\cite{PhysRevLett.92.076401}, coherent manipulation~\cite{PhysRevA.96.032342}, nanoscale sensing~\cite{Rondin_2014}. Among the most widely studied solid defects in diamond, nitrogen-vacancy~(NV) center~\cite{DOHERTY20131,doi:10.1063/PT.3.2549}, due to its bright and stable luminescence properties and long electron spin coherence time, is widely used in quantum computation~\cite{PhysRevA.105.012611,PhysRevA.100.052332,Zhou:15,PhysRevA.85.042306}, state detection~\cite{PhysRevLett.92.076401}, coherent manipulation~\cite{PhysRevA.96.032342}, nanoscale sensing applications~\cite{Rondin_2014}. However, the development of NV centers is limited by the characteristics of weak and unstable optical transitions~\cite{Bernien2013}. Recently, the negatively charged silicon-vacancy~(SiV$^-$)~(abbreviated as SiV) center has attracted great attention. The SiV center is formed by replacing two adjacent carbon atoms in the diamond lattice with a silicon atom, and the silicon atom is located between two vacancies. Due to its $D_{3d}$ point group symmetry that can protect it from optical inhomogeneity~\cite{doi:10.1126/science.aah6875,Dietrich_2014,https://doi.org/10.1002/pssa.201700586,PhysRevB.98.035306,PhysRevB.97.205444,PhysRevApplied.11.044022,PhysRevLett.128.203603,PhysRevLett.112.036405}, SiV center can realize practicable initialization and readout of qubits~\cite{PhysRevLett.113.263602,Pingault2017}. The scalability of SiV centers can be achieved by one-dimensional~(1D) phononic waveguide~\cite{PhysRevLett.120.213603,PhysRevA.105.032415} and photonic crystal cavity~\cite{Li2020}. In addition, the narrow width~(around 5 nm)~\cite{PhysRevX.8.021063}, the zero-phonon line emission~\cite{HAuBler_2017}, and optical coherence properties of SiV centers in diamond lay the foundation for quantum computation~\cite{Ladd2010}. So far, the SiV center has been very much in favor of different quantum information processes, including entanglement generation~\cite{Li2020,PhysRevA.101.042313}, spin-squeezed state preparation~\cite{PhysRevA.103.013709,https://doi.org/10.1002/qute.202000034}, topological phase simulation~\cite{PhysRevResearch.2.013121}, etc. Based on these studies, achieving quantum computing in SiV centers is highly anticipated.


Quantum computation based on the geometric phase relies on the global characteristics of the evolution, rather than the specific details of the evolution. The geometric phase has inherent noise-resilience features against certain local noises, making it a valuable resource for fault-tolerant quantum computation~\cite{PhysRevA.70.042316,PhysRevA.72.020301}. The development of the geometric phase has progressed from the initial adiabatic Abelian phase~(Berry phase)~\cite{doi:10.1098/rspa.1984.0023} to the adiabatic non-Abelian phase~\cite{PhysRevLett.52.2111}, and then to the nonadiabatic Abelian phase~\cite{ANANDAN1988171} and nonadiabatic non-Abelian phase~(Aharonov-Anandan phase)~\cite{PhysRevLett.58.1593}. Based on these phases, adiabatic Abelian geometric quantum computation~(GQC)~\cite{Jones2000}, adiabatic non-Abelian GQC~\cite{ZANARDI199994}, nonadiabatic Abelian GQC~(NGQC)~\cite{PhysRevLett.87.097901,PhysRevLett.89.097902}, and nonadiabatic non-Abelian GQC~\cite{Sj_qvist_2012,PhysRevLett.109.170501} have been proposed successively. In recent years, various optimization methods have been considered to combine with NGQC, such as optimal control~\cite{PhysRevA.102.042607,Yun:21,PhysRevA.103.062607}, time-optimal techniques~\cite{PhysRevApplied.10.054051,PhysRevApplied.14.064009}, shortened path methods~\cite{doi:10.1063/5.0071569,https://doi.org/10.1002/qute.202000140}, noncyclic shcemes~\cite{PhysRevResearch.2.043130}, reverse engineering scheme~\cite{PhysRevResearch.4.013233,PhysRevResearch.2.023295,Li_2021,PhysRevA.103.032609}, and so on~\cite{PhysRevLett.123.100501,PhysRevResearch.3.023104,PhysRevResearch.3.L032066,https://doi.org/10.1002/qute.201900013,https://doi.org/10.1002/andp.202100057,PhysRevA.103.032616}.  These methods have been demonstrated in different platforms, including superconducting circuits~\cite{PhysRevLett.124.230503,PhysRevLett.122.080501}, trapped ions~\cite{PhysRevLett.127.030502,PhysRevApplied.14.054062}, and NV centers in diamond~\cite{PhysRevApplied.16.024060,PhysRevLett.122.010503}, showing the robustness plasticity of GQC with respect to control error.

Given that a quantum system will inevitably be influenced by its environment, destroying the quantum information therein, various methods have been proposed to prevent quantum systems from destruction by environmental impact, including decoherence-free subspace~(DFS)~\cite{PhysRevLett.81.2594,PhysRevLett.95.130501,PhysRevLett.85.1762,doi:10.1126/science.290.5491.498,Lidar2003,PhysRevLett.85.1758,PhysRevLett.109.170501,PhysRevLett.97.140501}, noiseless subsystems~\cite{PhysRevLett.96.050501,doi:10.1126/science.1064460,PhysRevA.89.042302,PhysRevA.67.062303}, and dynamical decoupling~(DD)~\cite{2014,PhysRevLett.82.2417,doi:10.1098/rsta.2011.0355,PhysRevLett.95.180501,Biercuk2009}, etc. Among these methods, DD is attractive due to its low resource consumption and excellent performances~\cite{PhysRevA.90.022323,PhysRevA.103.012205,Wu2021,PhysRevA.102.032627,PhysRevA.93.022304,PhysRevA.91.042325,PhysRevA.86.050301,PRXQuantum.4.010334,PhysRevLett.82.2417,PhysRevLett.83.4888}. DD counteracts the interaction between the system and environment by using suitable external instantaneous intense pulse sequences, which can effectively improve the immunity of the quantum system to external environment. These rapid intense pulse sequences can be regarded as a generalization of spin-echo experiment~\cite{PhysRev.80.580} that approximately eliminated the effect of unwanted interaction. Therefore, combining NGQC with DD can protect quantum gates~\cite{PhysRevA.103.012205,PhysRevA.102.032627} by averaging out decoherence caused by the interaction between qubits and their environment, making it a valuable tool for quantum computation applications.

In this work, we propose to implement quantum computation with DD-protected nonadiabatic geometric gates in a hybrid system consisting of SiV centers placed in a 1D phononic diamond waveguide. The SiV centers are coupled to each other by strong strain, with a fixed distance between adjacent centers. To realize quantum gates, we encode the qubits in the long-lifetime ground states of the SiV centers, which are well-suited for this purpose.
To improve the system's immunity to the environment, we apply the DD technique by introducing a sequence of rapid pulses that eliminate the impact of the environment on the system. Our work presents an alternative approach to realizing NGQC in SiV centers, with several advantages. Firstly, SiV centers are stable and easy to operate, and their qubits are insensitive to spontaneous emission. Secondly, we use the geometric phase to realize quantum computation, which has built-in robustness against certain local noises. Finally, DD pulse sequences almost completely eliminate the impact of the environment on the system, making our scheme immune to decoherence caused by environmental factors.


\section{Physical model and effective Hamiltonian}
\label{siv}
\subsection{physical model and its Hamiltonian}

\begin{figure}[htbp]
	\centering \includegraphics[width=\linewidth]{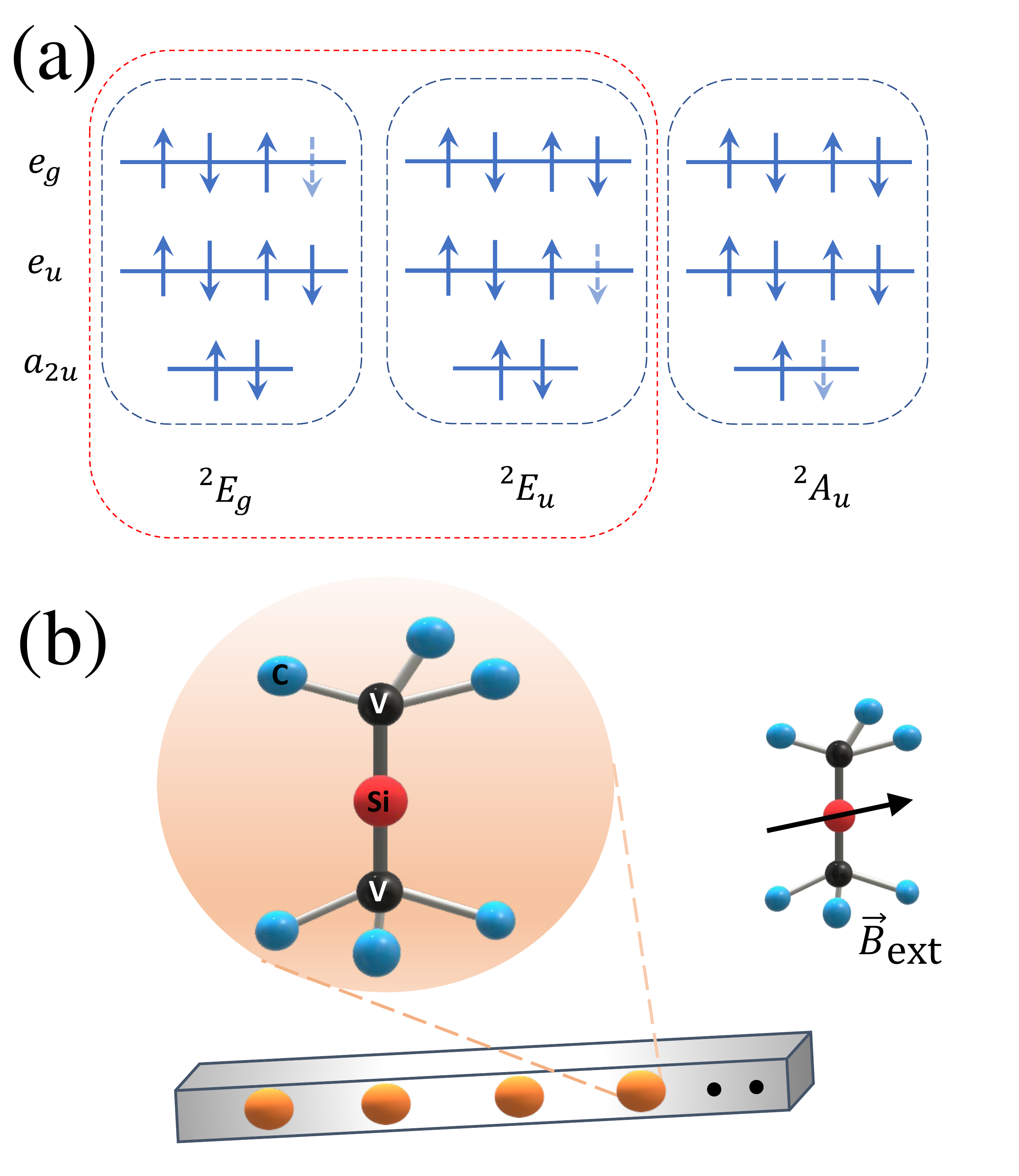}
	\caption{(a) SiV electronic structure. The blue arrows mean electrons, and the dashed arrows indicate electron holes. (b) Illustrative schematic. $N$ SiV centers  are placed in a phononic waveguide and the distance between two adjacent SiV centers is fixed. The right shows the direction of the external magnetic field.}
	\label{figlevel}
\end{figure}

Eleven electrons of the SiV result in the ground state $^2 E_g$ or excited states $^2 E_u$ and $^2 A_{2u}$ with a single unpaired electron, as shown in Fig.~\ref{figlevel}(a). In order to allow optical transitions between all levels, the magnetic field with $B=0.21 ~T$, and the direction between the magnetic and the [111] high-symmetry axis of the SiV is $70.5^\circ$~\cite{PhysRevLett.120.053603}. Then, the spin degeneracy is lifted. The structure of negatively charged SiV we consider is determined by the spin-orbit interaction, the Jahn-Teller~(JT) effect, and the Zeeman splittings. The Hamilton of the SiV center can be written as 
\begin{eqnarray}
\label{ham}
H_{\rm SiV}=-\lambda_{\rm SO}L_z S_z+H_{JT}+f\gamma_L \vec{B}\cdot \vec{L}+\gamma_s \vec{B}\cdot \vec{S},
\end{eqnarray}
where $\lambda_{\rm SO}$ represents the strength of the spin-orbit coupling. $\gamma_L$ and $\gamma_s$ are the orbital and spin gyromagnetic ratio, respectively. Suppose the external magnetic field is tilted from the positive direction along the $z$-axis, so the Zeeman splitting of Hamiltonian can be expressed as $\gamma_sB_zS_z+\gamma_sB_xS_x$. The strength of the JT effect coupling along $x$~$(y)$ can be denoted as $\Upsilon_x$~$(\Upsilon_y)$. Diagonalizing Eq.~(\ref{ham}), four ground eigenstates \{$|1\rangle,\ |2\rangle,\ |3\rangle,\ |4\rangle$\} and four excited eigenstates \{$|A\rangle,\ |B\rangle,\ |C\rangle,\ |D\rangle$\} can be obtained 
\begin{eqnarray}
|1\rangle & \equiv&|e_{g+}\searrow\rangle \approx|e_{g+}\downarrow\rangle-\eta_+|e_{g+}\uparrow\rangle,\ \ \notag\\
|2\rangle &\equiv &|e_{g-}\nearrow\rangle\approx |e_{g-}\uparrow\rangle-\eta_-|e_{g-}\downarrow\rangle,\ \ \notag\\
|3\rangle &\equiv&|e_{g-}\searrow\rangle \approx |e_{g-}\downarrow\rangle+\eta_-|e_{g-}\uparrow\rangle,\ \   \notag\\
|4\rangle &\equiv& |e_{g+}\nearrow\rangle\approx |e_{g+}\uparrow\rangle+\eta_+|e_{g+}\downarrow\rangle,\ \   \notag\\
|A\rangle & \equiv&|e_{u+}\searrow\rangle \approx|e_{u+}\downarrow\rangle-\eta_+|e_{u+}\uparrow\rangle,\ \ \notag\\
|B\rangle &\equiv &|e_{u-}\nearrow\rangle\approx |e_{u-}\uparrow\rangle-\eta_-|e_{u-}\downarrow\rangle,\ \ \notag\\
|C\rangle &\equiv&|e_{u-}\searrow\rangle \approx |e_{u-}\downarrow\rangle+\eta_-|e_{u-}\uparrow\rangle,\ \   \notag\\
|D\rangle &\equiv& |e_{u+}\nearrow\rangle\approx |e_{u+}\uparrow\rangle+\eta_+|e_{u+}\downarrow\rangle.\ \   \notag
\end{eqnarray}

\begin{figure}[htbp]
	\centering \includegraphics[width=\linewidth]{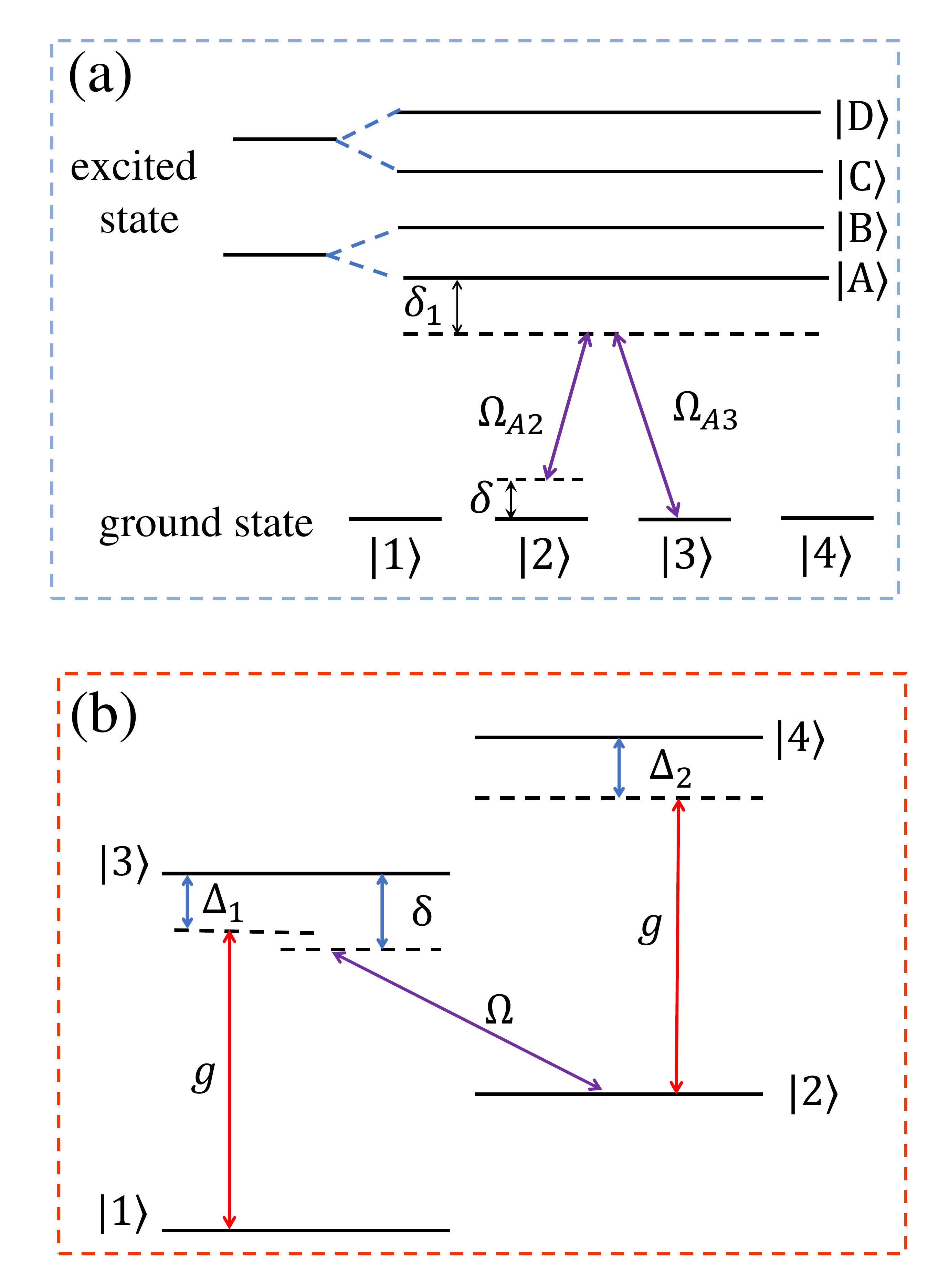}
	\caption{ (a) Energy level configuration of the $j$th SiV center. (b) Effective energy level structure.}
	\label{fig1}
\end{figure}

In the present system, only four ground states and the first excited state are considered. The Hamiltonian of a single SiV center can be expressed as 
\begin{eqnarray}
H_{\rm SiV} = \sum_{i=1}^4\omega_i |i\rangle |\langle i|+\omega_a|A\rangle\langle A|,
\end{eqnarray}
with $\omega_1\approx-\frac{\Delta+\omega_B}{2}-\frac{\eta_+\omega_x}{2},\ \omega_2\approx-\frac{\Delta-\omega_B}{2}-\frac{\eta_-\omega_x}{2},\ \omega_3\approx\frac{\Delta-\omega_B}{2}+\frac{\eta_+\omega_x}{2},\ \omega_4\approx\frac{\Delta+\omega_B}{2}+\frac{\eta_+\omega_x}{2},\ \eta_{\pm}=\frac{1}{2}\frac{\omega_x}{\Delta\pm\omega_B}\ $, and $\Delta=\sqrt{\lambda^2_{\rm SO}+4(\Upsilon_x^2+\Upsilon_y^2)}$. $|e_\pm\rangle$ are the eigenstates of the angular momentum operator and the up and down arrow denote the spin-up and spin-down state of the spin projections. $|e_{g(u)}\rangle$ means that the single unpaired electron appear at $e_{g(u)}$ level. We can find that each basis vector in the ground state subspace contains spin-up and spin-down components when the magnetic field and SiV axes are not oriented in the same direction, so the ground state basis vectors can be coupled to all energy levels.

 Two lasers coupling  $|2\rangle\leftrightarrow|A\rangle$ and  $|3\rangle\leftrightarrow|A\rangle$ with Rabi frequencies of $\Omega_{A2}$ and $\Omega_{A3}$ are used simultaneously, and the frequencies are $\omega_{A2}$ and $\omega_{A3}$, respectively, as illustrated in Fig.~\ref{fig1}~(a).
 The driving Hamiltonian can be written as 
\begin{eqnarray}
H_{\rm d}=\frac{\Omega_{A2}}{2}|A\rangle\langle2|e^{i\omega_{A2}t}+\frac{\Omega_{A3}}{2}|A\rangle\langle 3|e^{i\omega_{A3}t}+\rm H.c..
\end{eqnarray}
The detuning is $\delta_1=\omega_a-\omega_{A3}-\omega_3=\omega_a-\omega_{A2}-\delta$.

We consider $N$ SiV centers in a 1D phononic waveguide~[see Fig.~\ref{figlevel}~(b)], where the distance between two adjacent SiV centers is fixed, and the structure of the SiV center is shown in Fig.~\ref{fig1}~(a). For the phonon waveguide, the cross section $A$ is much larger than the length $L$, the phonon modes can be modeled as elastic waves with a displacement field $\vec{u}(\vec{r},t)$, obeying the equation of motion~\cite{PhysRevLett.120.213603},
\begin{eqnarray}
\rho \frac{\partial^2}{\partial t^2}\vec{u}=(\lambda+\mu)\vec{\nabla}(\vec{\nabla} \cdot \vec {u})+\mu \vec \nabla^2 \vec u
\end{eqnarray}
where the Lam\'e constants $\lambda=\frac{\nu E}{(1+\nu)(1-2\nu)}$, $\mu=\frac{E}{2(1+\nu)}$, $E$ is the Young's modulus, $\nu$ is the Poisson ratio, and $\rho $ is the mass density. The equation of motion meets the periodic boundary condition, and the amplitudes $A_{n,k}(t)$ obeys $\Ddot{A}_{n,k}(t)+\omega^2_{n,k}(t)A_{n,k}(t)=0$. The canonical coordinate $Q_{n,k}=(A_{n,k}+A^\ast_{n,-k})/\sqrt{2}$ and the canonical momenta $P_{n,k}$ can be written as 
\begin{eqnarray}
    Q_{n,k}=\sqrt{\frac{\hbar}{2M\omega_{n,k}}}(a^\dagger_{n,k}+a_{n,k}),\notag\\
     P_{n,k}=i \sqrt{\frac{\hbar M\omega_{n,k}}{2}}(a^\dagger_{n,k}+a_{n,k}),\notag
\end{eqnarray}
where $M=\rho A L$, and $a_{j,k}$ ($a^\dagger_{j,k}$) is the annihilation (creation) operator of the $k$th mode of the $j$th branch at frequency of $\omega_{j,k}$.

As for the strain coupling caused by the small displacement of the defect atoms, the interaction in the Born-Oppenheimer approximation~\cite{PhysRevB.97.205444,PhysRevB.94.214115} can be approximated as
\begin{eqnarray}
H_s=\sum_{n,j,k}g_{j,k,n}\hat{a}_{j,k}\hat{J}_+^n e^{ikx_n}+\rm H.c.,
\end{eqnarray}
where $\hat{J}_-=\hat{J}_+^\dagger=|1\rangle\langle3|+|2\rangle\langle4|$ means the lowering operator of the $k$th mode of the $j$th  branch of the $n$th SiV center, $g_{j,k,n}=d\sqrt{\frac{\hbar k^2 }{2\rho LA\omega_{j,k}}}\xi_{j,k}(y_n,z_n)$ is the coupling strength, $d/2\pi=1\ \rm PHz$ is the strain sensitivity, $\xi_{n,k}(y_n,z_n)$ is the dimensionless coupling profile accounting for the specific strain distribution and for a homogeneous compression mode $\xi(y,z)=1$, and $x_n$ denotes the position of the $n$th SiV center. So the Hamiltonian of phonon modes can be described by
\begin{eqnarray}
\hat{H}_{\rm ph}=\sum_{j,k} \omega_{j,k}\hat{a}^\dagger_{j,k}\hat{a}_{j,k}.
\end{eqnarray}
Then, the full Hamiltonian of the system is 
\begin{eqnarray}
H_{\rm full}=H_{\rm SiV}+H_{\rm ph} +H_d +H_s.
\label{eqfull}
\end{eqnarray}

\subsection{The effective Hamiltonian}

In the interaction picture, by considering the rotating-wave approximation and using effective Hamiltonian theory~\cite{doi:10.1139/p07-060,Brion_2007} with the condition of $\delta_1,\delta\gg \Omega_{A2},\Omega_{A3}$, the Hamiltonian of the $n$th SiV center can be rewritten as

\begin{eqnarray}
H_n(t)=&&ga(|3\rangle_n\langle 1|e^{i\Delta_1 t}+|4\rangle_n\langle 2|e^{i\Delta_2 t}) \notag\\
&&+\frac{\Omega}{2} |3\rangle_n\langle 2|e^{i\delta t} +\rm H.c.,\notag\\
\label{eq18}
\end{eqnarray} where $\Omega=-\Omega_{A2}^*\Omega_{A3}(2\delta_1+\delta)/ 4\delta_1(\delta_1+\delta)$~[see Fig.~\ref{fig1}(b)] and we have dropped the Stark shift that can be compensated by additional filed-induced energy shifts~\cite{PhysRevLett.97.083002,PhysRevLett.108.206401}.  

Further, in the condition of $\Delta_1, \Delta_2\gg g, \Omega$, the effective Hamiltonian can be denoted as 
\begin{eqnarray}
H_{n \rm eff}= \frac{\Omega_{\rm eff}}{2} a^\dagger |1\rangle_n \langle 2|e^{i(\delta-\Delta_1)t}+\rm H.c.
\label{eq7}
\end{eqnarray}
with $\Omega_{\rm eff}=\Omega g(\Delta_1+\delta)/2\Delta_1\delta$.
We can find that its form is consistent with the Jaynes-Cummings model that can be used to realize quantum computation~\cite{doi:10.1080/09500349314551321}.

\begin{figure}[htbp]
  \centering \includegraphics[width=\linewidth]{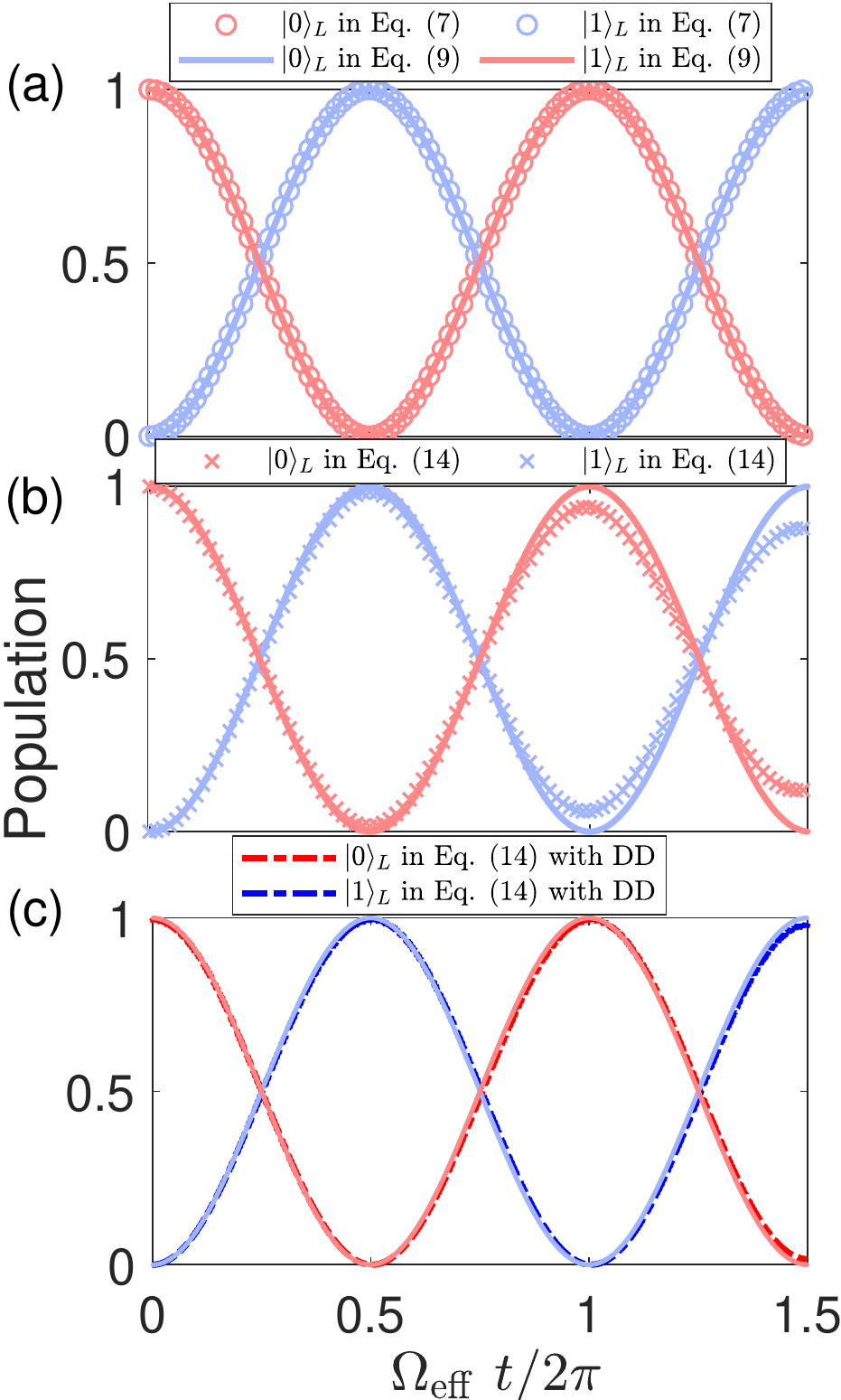}
	\caption{(a) Comparison of the population of $|0\rangle_L$ and $|1\rangle_L$ from Eq.~(\ref{eqfull}) with
that from Eq.~(\ref{eq7}). (b) (c) Comparison of the population of $|0\rangle_L$ and $|1\rangle_L$ from Eq.~(\ref{hsum}) with
that from Eq.~(\ref{eq7}) without (b) and with (c) DD-protected, respectively, where $\Omega/2\pi=10~\rm MHz$, $g/2\pi=5~\rm MHz$, $\Delta_1/2\pi= 100~\rm MHz$, $\Delta_2/2\pi=500~\rm MHz$, and $\delta=\Delta_1$.}
\label{fig2}
\end{figure}

Based on the effective Hamiltonian of the SiV center system in Eq.~(\ref{eq7}), we propose to conduct logic gates with SiV centers placed in the 1D phononic waveguide.
For simplicity, we consider the single excitation mode and regard $|11\rangle$ and $|20\rangle$ as logical qubits $|0\rangle_L$ and $|1\rangle_L$. $|11\rangle$~($|20\rangle$) is the abbreviation $|1\rangle\otimes|1\rangle$~($|2\rangle\otimes|0\rangle$) where the first ket in the product means the energy level of SiV center, and the second means the Fork state of phonon. The degree of conformity between the Hamiltonians in Eqs.~(\ref{eqfull}) and (\ref{eq7}) is shown in Fig.~\ref{fig2}(a). It can be seen that the effective Hamiltonian and the full Hamiltonian agree with each other very well. For the quantum system considered here, we modelize the interaction between qubits and environment as Hamiltonian   $H_I(t)=G_2[H(t)\otimes \sum _{k=x,y,z}\sigma_k]$, where $G_1/2\pi=G_2/2\pi=40~\rm kHz$.
 From Fig.~\ref{fig2}(b), we can see that with the increase of time, the environment has an increasingly significant impact on the system. Fortunately, after applying DD sequences XY-4, the effect of environment can be  eliminated approximately, as shown in Fig.~\ref{fig2}(c), which proved that DD is an effective method for protecting the system coherence. 

\begin{figure}[htbp]
	\centering \includegraphics[width=\linewidth]{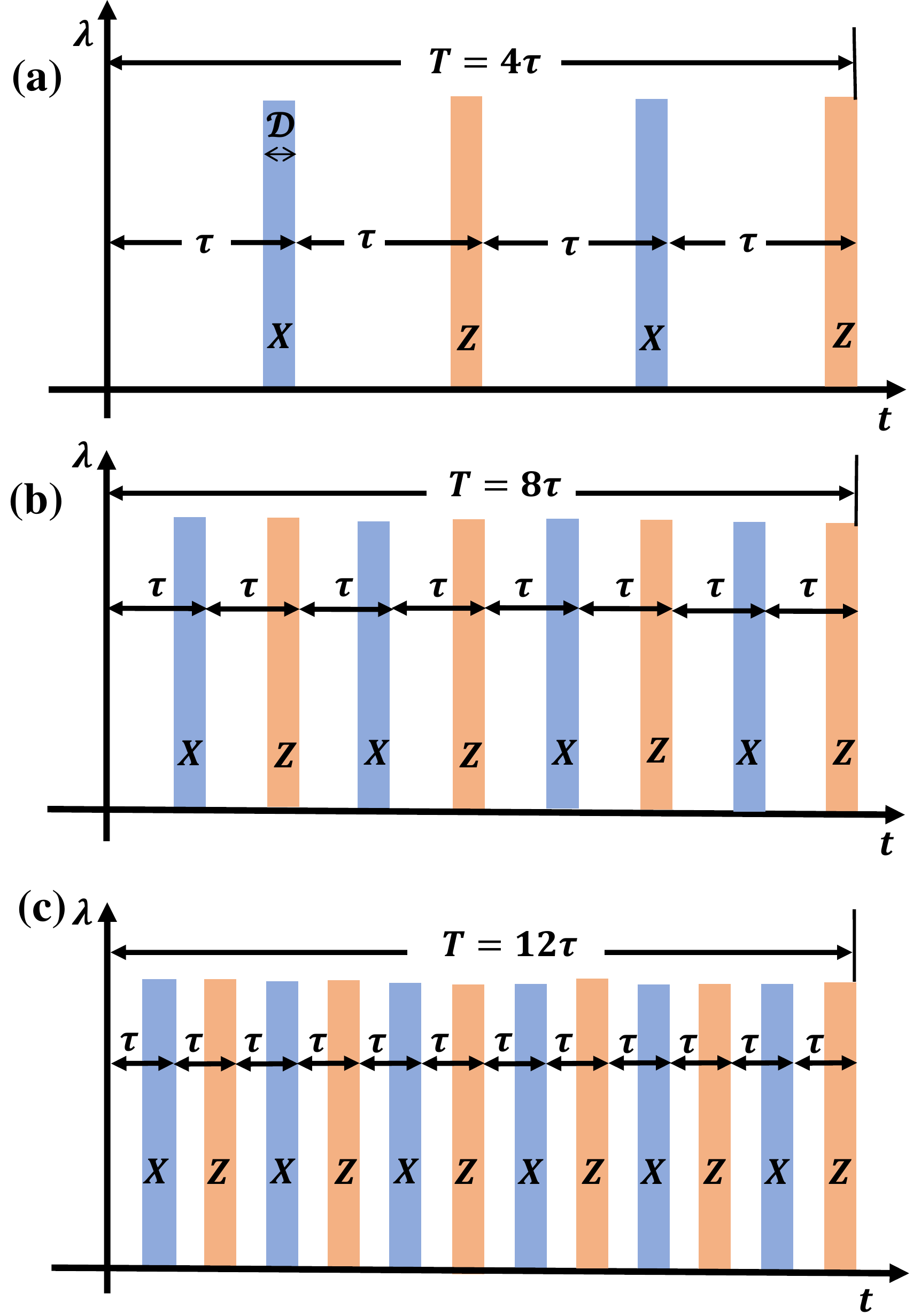}
	\caption{Schematic diagram of periodic decoupled pulses. (a) The XY-4 sequences. (b) The XY-8 sequences. (c) The XY-12 sequences.}
\label{pluse}
\end{figure}

\begin{figure}[htbp]
	\centering \includegraphics[width=\linewidth]{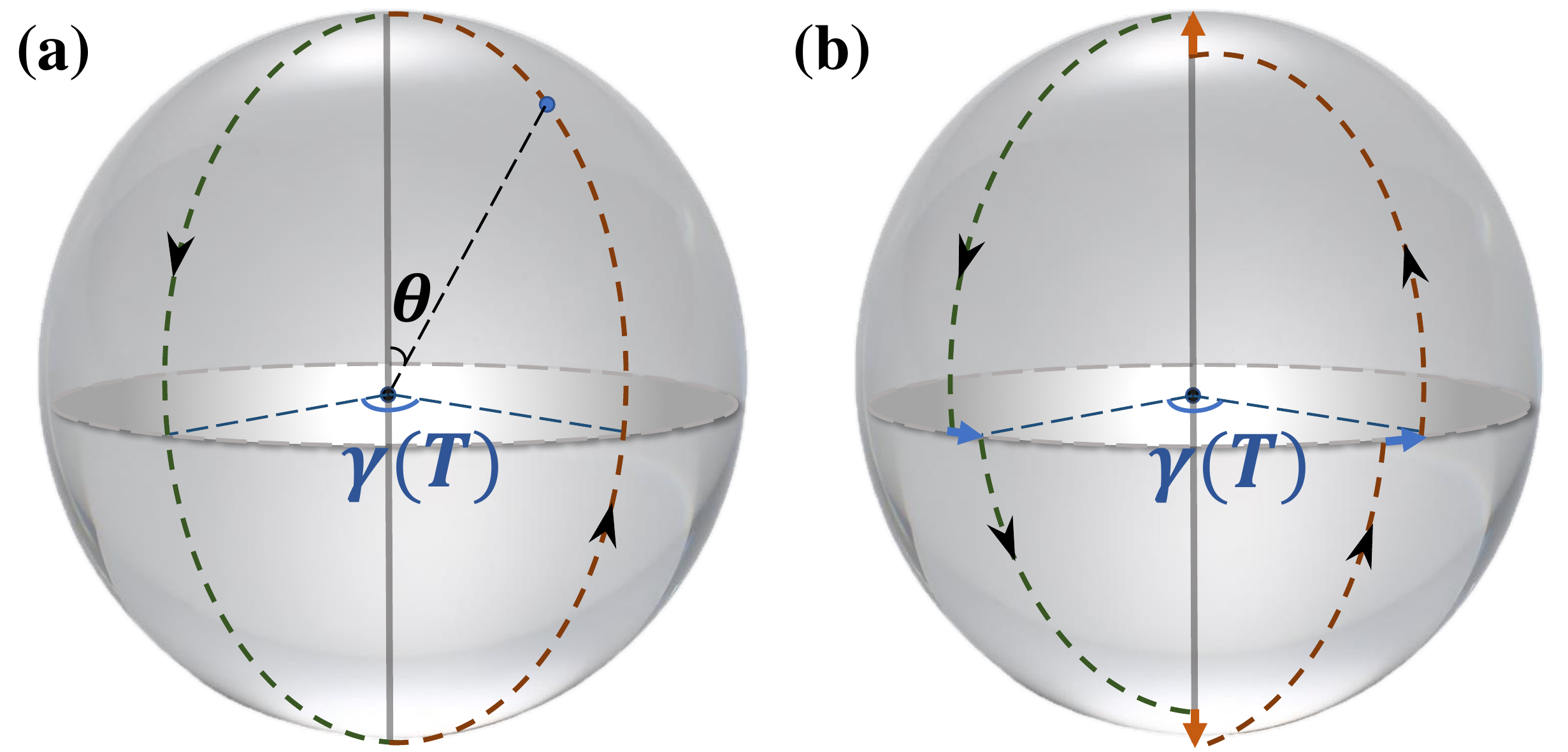}
	\caption{Evolution paths for the single-loop NGQC scheme (a) without and (b) with impact of the environment and decoupled pulses~(take the basic pulse as an example).}
\label{bloch}
\end{figure}

Inspired by this, we consider adding the decoupling sequence XY-~4 sequence and periodic DD sequences shown in Fig.~\ref{pluse} to the system to further improve the performance of the gate. Multiple fast and strong pulses with pulse area being $\pi/2$ are applied, which can remove the impact of environment to a great extent~(details are presented in appendix \ref{a2}). The schematic diagram of the evolution trajectory with impact of the environment and decoupled pulses is shown in Fig.~\ref{bloch}(b). Because of the effect of environment, the trajectory of the evolution state will have a slight deviation from the ideal situation shown in Fig.~\ref{bloch}(a). This deviation can be eliminated by the DD pulses (represented by orange and blue arrows with solid lines), so as to achieve the ideal evolution. Consequently, the decoherence caused by the interaction between the quantum system and environment can be inhibited greatly.

\section{Implementation}

By combining a set of universal single-qubit gates and a nontrivial two-qubit gate, universal quantum computation can be realized~\cite{PhysRevLett.89.247902}. In this section, we show how to use SiV centers to realize general DD-protected nonadiabatic geometric quantum logic gates.
\subsection{Single-qubit gates}
\label{sec4b}

To construct general single-qubit gates, we choose a pair of orthogonal auxiliary bases 

\begin{eqnarray}
\begin{split}
 |\varphi_1(t)\rangle&=\sin{\frac{\theta(t)}{2}}e^{-i\phi(t)}|0\rangle_L-\cos{\frac{\theta(t)}{2}}|1\rangle_L,&\\
|\varphi_2(t)\rangle&=\cos{\frac{\theta(t)}{2}}|0\rangle_L+\sin{\frac{\theta(t)}{2}e^{i\phi(t)}}|1\rangle_L.&
\end{split}
\end{eqnarray}
Then, we have  $\Omega_{\rm eff }(t)=i e^{i\phi(t)} [\dot{\theta}(t)+i\cos{\theta}(t)\sin{\theta(t)}\dot{\phi}(t)]$ and $\Delta_{\rm eff}(t)=-\frac{1}{2}\sin^2{\theta(t)}\dot{\phi}(t)$, where $\Delta_{\rm eff}=\delta-\Delta_1$~( details are presented in appendix \ref{a1}).

We take a single qubit phase gate and a NOT gate as  examples. For the phase gate, the polar angle~$\theta(t)$ and the azimuth angle~$\phi(t)$ varies by starting from the north pole $[\theta(0)=0, \phi(0)]$, passing through the south pole, experiencing a sudden change~$[\theta(t)=\pi, \ \phi(t)=\phi(0)+\pi/4]$, and finally returning to the north pole. 
The Hamiltonian can be expressed as
\begin{align}
\begin{split}
\left \{
\begin{array}{ll}
\frac{\dot{\theta}(t)}{2}[-\sin{\phi(0)\sigma_x}+\cos{\phi(0)}\sigma_y],                    & t\in [0, \frac{T}{2}],\\
    \frac{\dot{\theta}(t)}{2}[-\sin({\phi(0)+\frac{\pi}{4})\sigma_x}+\cos({\phi(0)}+\frac{\pi}{4})\sigma_y],     & t\in (\frac{T}{2}, T],
\end{array}
\right.
\end{split}
\label{eq19}
\end{align}
and the parameters meet $\int_0^{T/2}\dot{\theta}(t)\rm dt=\pi/2$. $\sigma_x(y)$ is the Pauli matrix, and $\int_{T/2}^{T}\dot{\theta}(t)\rm dt=-\pi/2$. The schematic diagram of the evolution path is shown in Fig.~\ref{bloch}(a).

For the NOT gate~(ignore the global phase), the polar angle and the azimuth angle start from $[\theta(0)=\frac{\pi}{2},\  \phi(0)=0]$, passing the south pole and the north pole and back to the starting point. The parameters in the Hamiltonian satisfy
\begin{align}
\left \{
\begin{array}{ll}
\int_0^{T_1} \dot{\theta}(t) dt=\theta(0), \ \ \ \ \phi(t)=0,  & t\in [0, T_1],\\[2mm]
 \int_{T_1}^{T_2} \dot{\theta}(t) dt=\pi, \ \ \ \ \phi(t)=-\frac{\pi}{2}  ,     & t\in (T_1, T_2],\\[2mm]
\int_{T_2}^{T} \dot{\theta}(t) dt=\pi-\theta(0), \ \ \  \phi(t)=0 ,     & t\in (T_2, T].
\end{array}
\right.
\label{eqH}
\end{align}

\begin{figure}[htbp]
	\centering \includegraphics[width=\linewidth]{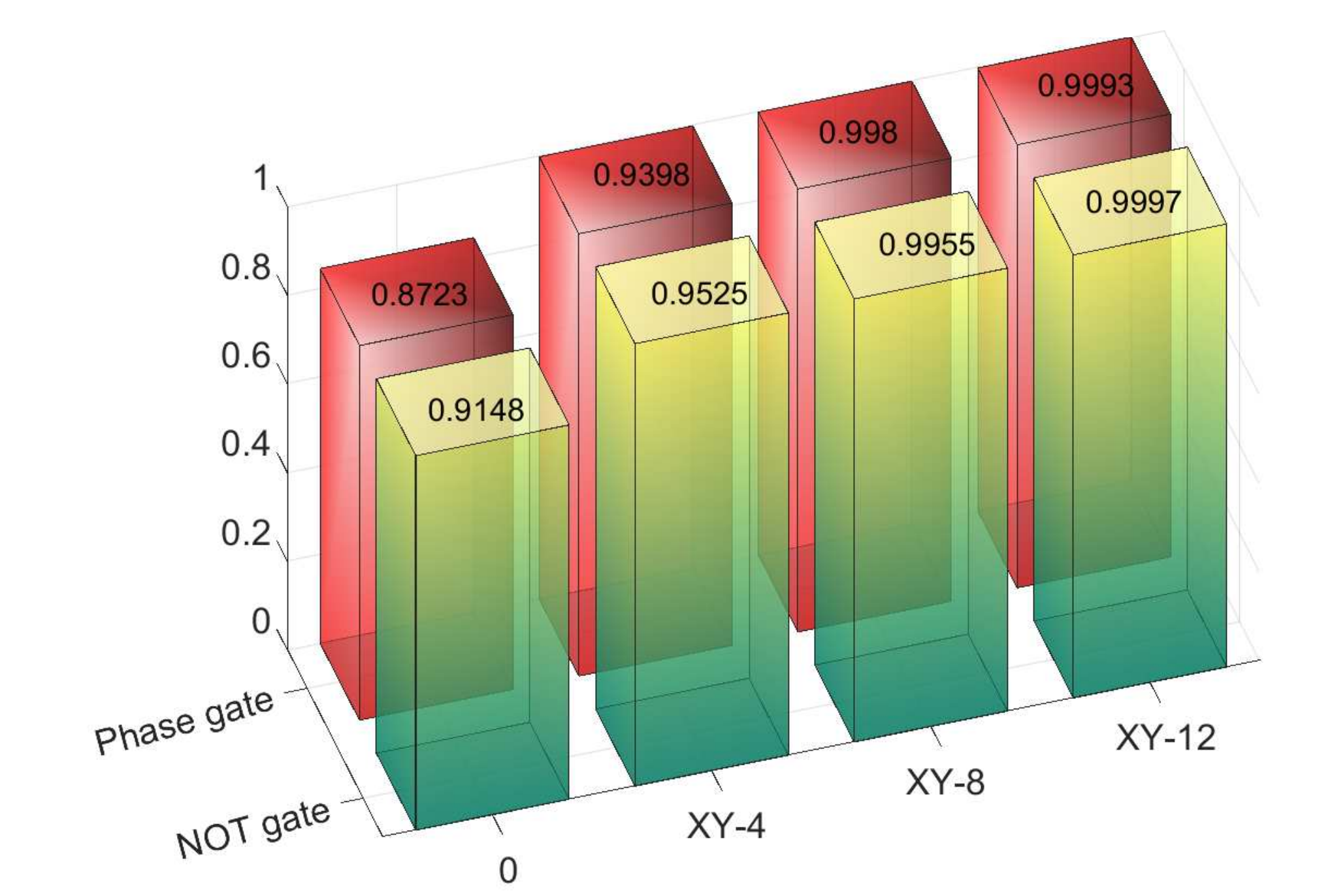}
	\caption{ Fidelity of the phase gate and the NOT gate based on the conditions in Eq.~(\ref{eq19}) and Eq.~(\ref{eqH}) without DD and with XY-~4 sequences, XY-~8 sequences, and XY-~12 sequences.  The control parameters are set the same as Fig.~\ref{fig2}(a).}
	\label{figfidelity}
\end{figure}

We use the fidelity defined as 
\begin{eqnarray}
F=\langle \psi_{\rm ideal}|\rho(T)|\psi_{\rm ideal}\rangle
\label{fidelity}
\end{eqnarray}
to evaluate the performance of the gate, where $|\psi_{\rm ideal}\rangle =U(T)|\psi(0)\rangle$ is the ideal state, $\rho$ is the density matrix. According to Ref.~\cite{PhysRevLett.105.230503}, the Liouville equation $\dot{\rho}_{\rm sum}(t)=-i[H_{\rm sum}(t),\rho_{\rm sum}(t)]$ can be used to calculate the density matrix, where 
\begin{eqnarray}
    H_{\rm sum}(t)=H(t)+H_E(t)+H_I(t)
    \label{hsum}
\end{eqnarray} is the total Hamiltonian, $H_E(t)$ is the Hamiltonian of the environment, and $H_I(t)$ is the Hamiltonian of interaction between the system and environment. The density operator $\rho(t) =\rm Tr_{E} \rho_{\rm sum}(t)$. In the numerical simulations, we use the single-qubit NOT (phase) gate, and the evolution operator is described as $U(T)=\rm exp(-i\pi \sigma_x /2)$ [$U(T)=\rm exp(-i\pi \sigma_z /2)$]. The initial state is selected as $|\psi (0)\rangle=|0\rangle_L$ for simplicity, and the environment Hamiltonian is $H_E(t) = G_1(I_{\rm sys}\otimes \sum _{k=x,y,z}\sigma_k)$ with $I_{\rm sys} $ meaning the identity matrix of the system, the $H_I(t)=G_2[H(t)\otimes \sum _{k=x,y,z}\sigma_k]$. From Fig.~\ref{figfidelity}, we can see the fidelity of the NOT (phase) gate is significantly improved, where when the number of decoupling pulse sequences are XY-~4, XY-~8, and XY-~12, the fidelities can reach 0.9525~(0.9398), 0.9955~(0.998), and 0.9997~(0.9993), respectively.
 
\begin{figure}[htbp]
	\centering \includegraphics[width=\linewidth]{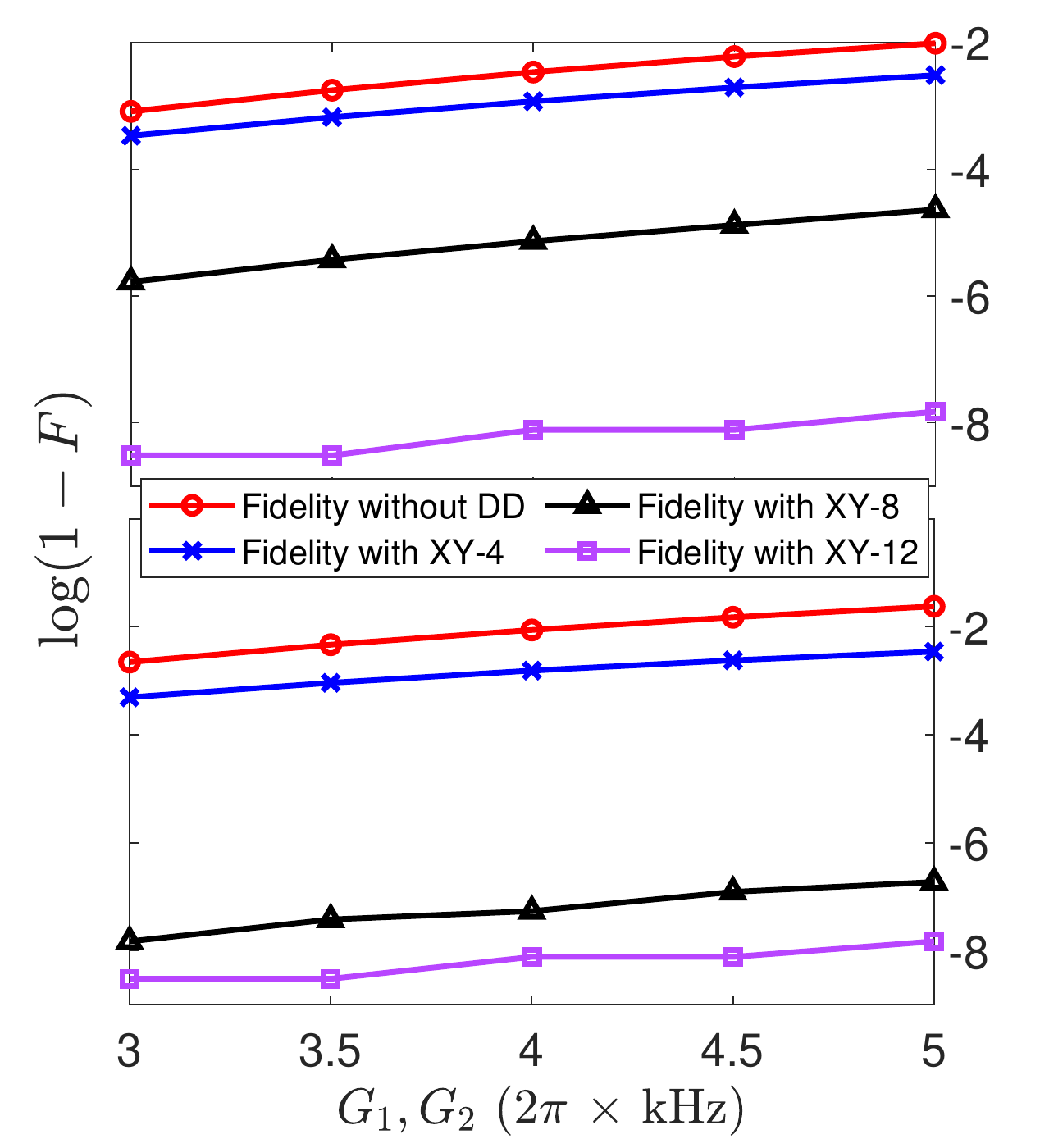}
   \caption{The robustness of (a) the NOT gate and (b) the phase gate varying with the strength of environmental interaction. The control parameters are set the same as Fig.~\ref{fig2}.}
\label{figrobust}
\end{figure}

We further demonstrate the robustness of our scheme against the decoherence and dephasing with coefficients $G_1$ and $G_2$, respectively. The robustness of the NOT gate~(phase gate) changing with the strength of the environmental interaction is shown in Fig.~\ref{figrobust}(a)[(b)]. From this, we can see that the DD greatly protects the nonadiabatic geometric quantum gates from environment-induced decoherence and decay.

\subsection{Two-qubit quantum gate}
\label{sec4c}
Here, we consider two SiV centers in the phononic waveguide, and define the effective detuning  $\delta-\Delta_1=\Lambda_1$. The Hamiltonian becomes
\begin{eqnarray}
    H_n=\frac{\Omega_{\rm eff}}{2}a^\dagger|1\rangle\langle2|e^{i\Lambda_n t},
    \label{eq22}
\end{eqnarray}
for the $n$th SiV center coupled to the phononic waveguide. When $\Lambda_n\gg \Omega_{\rm eff}$, the effective Hamiltonian of the system can be written as
\begin{eqnarray}
H_{\rm two}=-\Omega_{\rm eff}^2\frac{\Lambda_1+\Lambda_2}{8\Lambda_1\Lambda_2}\sigma_1^-\sigma_2^+ e^{i (\Lambda_1-\Lambda_2) t}+\rm H.c.,
\label{eq23}
\end{eqnarray}
where $\sigma_1^-$~($\sigma_2^+$) denotes the energy level decline~(rise) operator of the first~(second) SiV center. Now, similar to the single qubit gate, we encode $|1\rangle_{\rm SiV}$~($|2\rangle_{\rm SiV}$) as logical qubit $|0\rangle_L$~($|1\rangle_L$), so the Hamiltonian in Eq.~(\ref{eq23}) can be written in the following form in the basis space \{$|00\rangle_L\ |01\rangle_L\ |10\rangle_L\ |11\rangle_L $\},
\begin{eqnarray}
H_2=-
\begin{pmatrix}
    0&0&0&0\\
    0&\Lambda_2-\Lambda_1&\frac{\Omega^2_{\rm eff}(\Lambda_1+\Lambda_2)}{8\Lambda_1 \Lambda_2}&0\\
    0&\frac{\Omega^2_{\rm eff}(\Lambda_1+\Lambda_2)}{8\Lambda_1 \Lambda_2}&\Lambda_1-\Lambda_2&0\\
    0&0&0&0\\
\end{pmatrix}.
\end{eqnarray}
Similar to the single-qubit case, we choose four auxiliary bases
\begin{eqnarray}
&&|\varphi_1(t)\rangle=|00\rangle_L,\notag\\
&&|\varphi_2(t)\rangle=\cos{\frac{\theta(t)}{2}}|01\rangle_L+\sin{\frac{\theta(t)}{2}e^{i\phi(t)}}|10\rangle_L,\notag\\
&&|\varphi_3(t)\rangle=\sin{\frac{\theta(t)}{2}}e^{-i\phi(t)}|01\rangle_L-\cos{\frac{\theta(t)}{2}}|10\rangle_L,\notag\\
&&|\varphi_4(t)\rangle=|11\rangle_L.
\end{eqnarray}
Then, we have $\mathop{O}_{\rm eff}\equiv \Omega^2_{\rm eff}\frac{\Lambda_1+\Lambda_2}{4\Lambda_1\Lambda_2}=i e^{i\phi(t)} [\dot{\theta}(t)+i\cos{\theta}(t)\sin{\theta(t)}\dot{\phi}(t)]$ and $\Lambda_{\rm eff}(t)=\frac{1}{2}\sin^2{\theta(t)}\dot{\phi}(t)$, where $\Lambda_{\rm eff} = \Lambda_2-\Lambda_1 $~(details are presented in appendix~\ref{a1}). From Fig.~\ref{fig2}(b) we know that the full Hamiltonian and the effective Hamiltonian are exactly consistent.

Here, we choose $\Lambda_1=\Lambda_2$, $\gamma(T)=\pi/2$, $\theta(0)=\pi/2$, and $\varphi(0)=\pi$, the parameters in Hamiltonian meet 
\begin{align}
\left \{
\begin{array}{ll}
\int_0^{T_1} \mathop{O}_{\rm eff} dt=\theta(0), \ \ \ \ \phi(t)=\pi,  & t\in [0, T_1],\\[2mm]
 \int_{T_1}^{T_2} \mathop{O}_{\rm eff} dt=\pi, \ \ \ \ \phi(t)=\pi+\frac{\pi}{2}  ,     & t\in (T_1, T_2],\\[2mm]
\int_{T_2}^{T} \mathop{O}_{\rm eff} dt=\pi-\theta(0), \ \ \  \phi(t)=\pi  ,     & t\in (T_2, T],
\end{array}
\right.
\label{eq26}
\end{align} 
the azimuth angle $\phi$ changes twice at the south pole and north pole, an iSWAP gate can be achieved, which is a universal gate for quantum computation\cite{PhysRevA.67.032301}.

The system Hamiltonian is $H_2(t)$ in Eq.~(\ref{eq18}), the environment Hamiltonian is represented as $H_E(t)=G_1(I_{\rm sys}\otimes \sum _{k=x,y,z}\sigma_k)$, and interaction Hamiltonian is described as $H_I(t)=G_2[H_2(t)\otimes \sum _{k=x,y,z}\sigma_k]$, where $G_1/2\pi=G_2/2\pi=1~\rm kHz$. Parameters are the same as Sec.~\ref{sec4b}.

\begin{figure}[htbp]
	\centering \includegraphics[width=\linewidth]{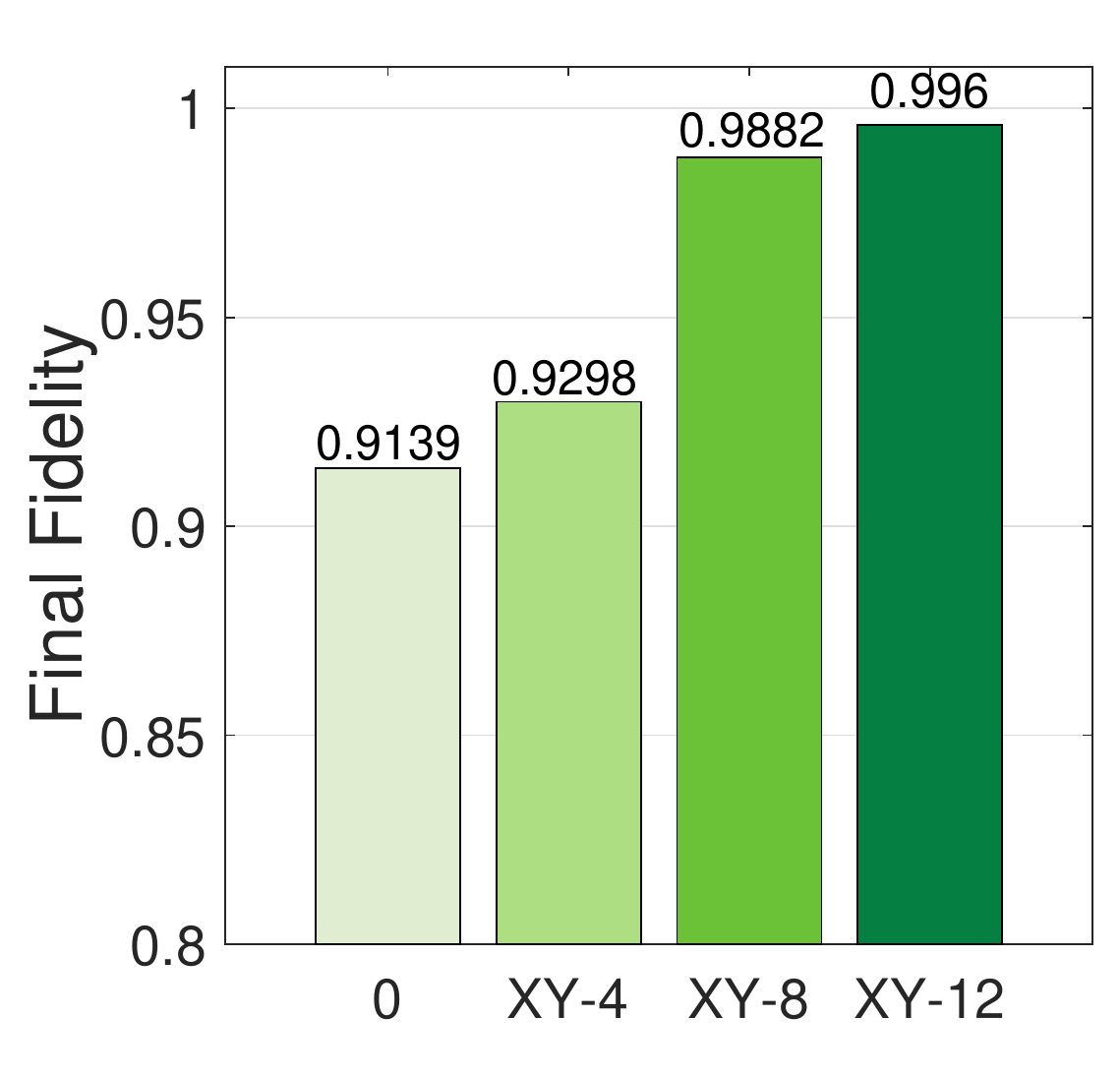}
	\caption{Final fidelity of the proposed iSWAP gate in SiV center based on the conditions in Eq.~(\ref{eq26}) without DD and with XY-~4 sequences, XY-~8 sequences, and XY-~12 sequences. The control parameters are set the same as Fig.~\ref{fig2}(b).}
\label{fig9}
\end{figure}

\begin{figure}[htbp]
	\centering \includegraphics[width=\linewidth]{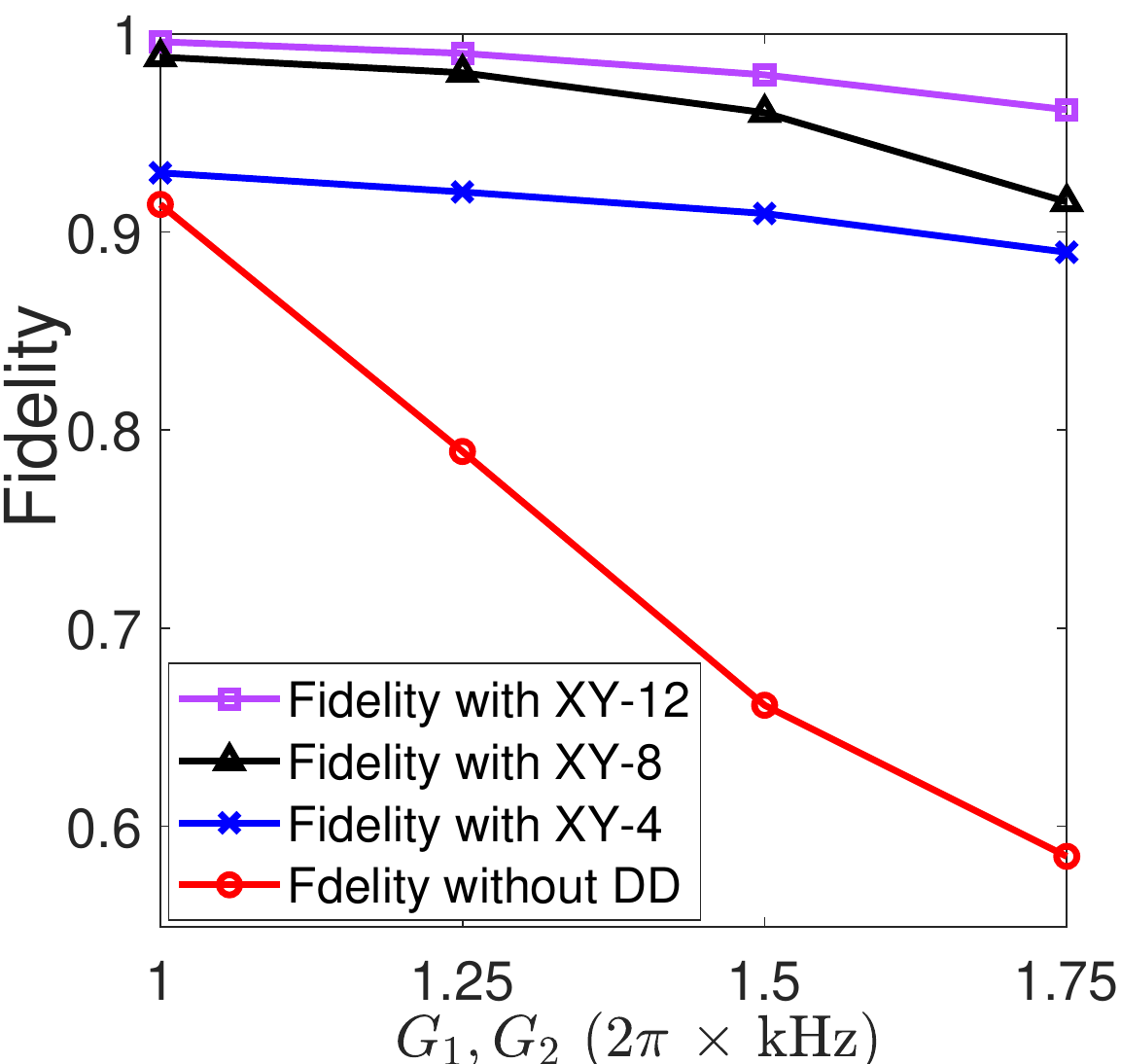}
	\caption{Fidelity of the proposed iSWAP gate in SiV center varying with $G_1$ and $G_2$. The control parameters are set the same as Fig.~\ref{fig9}.}
\label{fig10}
\end{figure}

The initial state is set as $|01\rangle_L $ for simplicity, and the fidelity $F$ defined as Eq.~(\ref{fidelity}) is shown as Fig.~\ref{fig9}. Due to environmental impact, the decoherence and dephasing coefficients ($G_1$ and $G_2$) compared with the effective Rabi frequency $\mathcal{O}_{\rm eff}$ of the model are too large to maintain a high fidelity, as shown in Fig.~\ref{fig9}(a). Further, we introduce the DD sequences shown in Fig.~\ref{pluse} to further improve the performance of the gate. When the number of DD pulse sequences is four, eight, and twelve, the fidelity can reach 0.9298, 0.9882, and 0.996, respectively. From the result of numerical simulations, one can see that the DD technique can protect the quantum system from decoherence and dephasing coming from the environment.
The fidelity of the iSWAP gate varying with $G_1$ and $G_2$ is shown in Fig.~\ref{fig10}, which shows the robustness of DD sequences.

\section{Experimental feasibility}
\label{sec5}
In our scheme, the considered SiV centers are coupled by a phononic waveguide. For the state-of-the-art techniques, SiV center arrays can be assembled successfully~\cite{PhysRevApplied.7.064021}, so as to implement the experimental structure. For the phonon waveguide, one can assume that the length is $L=80\ \rm \mu m$, the cross section $A=80\times80 \ \rm nm^2$, the Young's modulus $E=1050\ \rm GPa$, the Poisson ratio $\nu=0.2$, the mass density $\rho=3500\ \rm kg/m^3$~\cite{PhysRevLett.120.213603}, and the resulting coupling strength $g/2\pi\approx  5~\rm MHz$. The condition that the single excitation mode of phonon can be satisfied when the temperature $T$ is below millikelvin~\cite{PhysRevLett.120.053603}. Therefore, the relevant parameters of our scheme could be feasible in experiment. 

\section{conclusion }
In conclusion, we have proposed to implement quantum computation in the system of SiV centers for the first time by conducting universal nonadiabatic geometric quantum gates, and the dynamical decoupling pulse sequences are used to improve the immunity of the quantum system to environmental impact. The scalability can be achieved by a 1D phononic waveguide. Besides the scalability, our scheme is easy to initialize and read out. In addition, logical states of qubits are encoded in long-lifetime ground states of SiV centers, which can reduce the effect of decoherence caused by spontaneous emission. In summary, our scheme can protect quantum gates from both the control error caused by actual operation due to the intrinsic robustness of geometric quantum logic gates and the dephasing caused by the environment, which are two obstacles to the realization of high-fidelity quantum logic gates. Therefore, our scheme may provide a significant reference and pave an alternative path for implementing high-fidelity geometric quantum computation in the solid-state system.


 \section{Acknowledgement}
This work was supported by the National Natural Science Foundation of China under Grants (No. 12274376, No. U21A20434, No. 12074346), and a major science and technology project of Henan Province under Grant No. 221100210400, and the Natural Science Foundation of Henan Province under Grant No. 232300421075 and 212300410085, and Cross-disciplinary Innovative Research Group Project of Henan Province under Grant No. 232300421004.

\appendix

\section{Universal dynamical decoupling pulse sequences}\label{a2}

In the process of gate implementation, evolution inevitably receives influence from the surrounding environment, which can lead to a reduction of fidelity. In this appendix, a set of general pulse sequences that can cancel the system and environment are derived.

We consider the system-environment coupling Hamiltonian~\cite{2014,doi:10.1098/rsta.2011.0355,PhysRevLett.95.180501}
 \begin{eqnarray}
 H_{\rm se}=H_{\rm s}\otimes H_{\rm e}=\sum_{\alpha=x,y,z}\sigma^\alpha \otimes B^\alpha,
 \end{eqnarray}
 where $\sigma^\alpha $ denotes the system Hamiltonian and $B^\alpha$ expresses the environment operator.
The system evolution operator expressed as $f_\tau=e^{-i\tau H_{\rm se}}$.
Then, we apply several pluses to the system. We assume these pulses last for a period of time $\mathcal{D}$ with strength $\lambda$, and $\mathcal{D}\lambda=\frac{\pi}{2}$. When $\mathcal{D}\rightarrow 0$, $\lambda\rightarrow \infty$, the condition $\mathcal{D}\lambda=\frac{\pi}{2}$ can be satisfied. 

Now, we apply two X-pulse $X=e^{i\mathcal{D}\lambda\sigma^x}\otimes  I_B=e^{-i\frac{\pi}{2}\sigma^x}\otimes I_B=-i\sigma^x\otimes I_B$ on the system. Then, the evolution operator can be expressed as
\begin{equation}
\sigma^xH_{\rm se}\sigma^{x}=\sigma^x\otimes B^x -\sigma^y\otimes B^y-\sigma^z\otimes B^z,
\end{equation}
which means that $Xf_\tau Xf_\tau$ pluse sequence can cancel both the $y$ and $z$ contributions from environment. The evolution operator can be described as 
\begin{small}
\begin{eqnarray}
Xf^\prime_{2\tau}=Xf_\tau Xf_\tau=e^{-2i\tau(\sigma^x\otimes B^x+H_e)}+\mathcal{O}(\tau^2).
\end{eqnarray}
\end{small}
Then we apply Y-pulse $Y=e^{i\mathcal{D}\lambda\sigma^y}\otimes  I_B=e^{-i\frac{\pi}{2}\sigma^y}\otimes I_B=-i\sigma^y\otimes I_B$ on the system, 
\begin{small}
	\begin{eqnarray}
	Yf^\prime_{2\tau} Yf^\prime_{2\tau}&=&YXf_\tau Xf_\tau YX f_\tau Xf_\tau\\\notag
	&=&Zf_\tau Xf_\tau Zf_\tau Xf_\tau\\\notag
	&=&e^{-i4\tau H_e}+\mathcal{O}(\tau^2),
	\end{eqnarray}
\end{small}
 so evolution time $t=4\tau$, the system can decouple from the environment, i.e., $ZXZX$ pulse sequences are universal decoupling pulse sequences that can reduce the impact of environment on the system. Here, we can repeat the basic DD sequence periodically named periodic dynamical decoupling~\cite{PhysRevA.90.022323} to further enhance the performance of the system. The basic pulse sequence named XY-4 sequences, the pulse sequence with a period of two named XY-8 sequences, and the pulse sequence with a period of three named XY-12 sequences are shown in Fig.~\ref{pluse} (a), (b), and (c), respectively.

\section{Nonadiabatic geometric Hamiltonian}\label{a1}
This appendix will use the reverse-engineering scheme to give the target Hamiltonian for a system satisfying nonadiabatic geometric conditions~\cite{PhysRevResearch.2.023295}.
 For the single-qubit case, the orthogonal auxiliary bases can be chosen as
\begin{eqnarray}
\begin{split}
 |\varphi_1(t)\rangle&=\sin{\frac{\theta(t)}{2}}e^{-i\phi(t)}|0\rangle-\cos{\frac{\theta(t)}{2}}|1\rangle,&\\
|\varphi_2(t)\rangle&=\cos{\frac{\theta(t)}{2}}|0\rangle+\sin{\frac{\theta(t)}{2}e^{i\phi(t)}}|1\rangle,&
\end{split}
\end{eqnarray}
where $|0\rangle$ and $|1\rangle$ are two logical states of qubit, and $\theta(t)$ and $\phi(t)$ are the time-dependent parameters.
Then, the Hamiltonian of the system can be expressed as
\begin{eqnarray}
\label{eq2}
H(t)&=&i\sum_{k\neq l}^{2}\langle \varphi_l(t)|\dot{\varphi_k}(t)\rangle|\varphi_l(t)\rangle\langle\varphi_k(t)|\notag\\
&=&\Delta(t)(|1\rangle\langle1|-|0\rangle\langle0|)+[\frac{\Omega(t)}{2}|1\rangle\langle0|+\rm H.c.].\notag\\
\end{eqnarray}
Then, we have $\Omega(t)=i e^{i\phi(t)} [\dot{\theta}(t)+i\cos{\theta}(t)\sin{\theta(t)}\dot{\phi}(t)]$ and $\Delta(t)=-\frac{1}{2}\sin^2{\theta(t)}\dot{\phi}(t)$.

Corresponding to the auxiliary basis vectors, the initial states of the system are
\begin{eqnarray}
\begin{split}
|\psi_1(0)\rangle&= |\varphi_1(0)\rangle=\sin{\frac{\theta(0)}{2}}e^{-i\phi(0)}|0\rangle-\cos{\frac{\theta(0)}{2}}|1\rangle,&\\
     |\psi_2(0)\rangle&=|\varphi_2(0)\rangle=\cos{\frac{\theta(0)}{2}}|0\rangle+\sin{\frac{\theta(0)}{2}e^{i\phi(0)}}|1\rangle.
\end{split}
\end{eqnarray}

After a cyclic evolution, the evolution operator can be written as
\begin{widetext}
\begin{eqnarray}
U(T)&=&e^{-i\gamma(T)}|\psi_1(0)\rangle\langle\psi_1(0)|+e^{i\gamma(T)}|\psi_2(0)\rangle\langle\psi_2(0)|\notag\\
&=&\left(
\begin{array}{cc}
\cos{\gamma(T)}+i \cos{\theta_0}\sin{\gamma(T)}                           &ie^{-i\phi_0}\sin{\gamma(T)}\sin{\theta_0}    \\
ie^{i\phi_0}\sin{\gamma(T)}\sin{\theta_0}                &\cos{\gamma(T)}-i \cos{\theta_0}\sin{\gamma(T)}
\end{array}
\right)\nonumber\\
&=&e^{-i\gamma(T)\bf n\cdot \boldsymbol{\sigma}},
\end{eqnarray}
\end{widetext}
where $\theta_0\equiv \theta(0)$, $\phi_0\equiv \phi(0)$, $\gamma(T)=i \int_0^T\langle \varphi_1(t)|\dot{\varphi_1}(t)\rm dt$$= -i \int_0^T\langle \varphi_2(t)|\dot{\varphi_2}(t)\rm dt$ $=\frac{1}{2}\int_0^T [1-\cos\theta(t)\dot{\phi}(t)]\rm dt$ $=\frac{1}{2}\oint_C [1-\cos\theta(t)] d\phi$. It can be clearly seen that $\gamma(T)$ is a half of the solid angle, independent of the evolution details, and $\bf{n}=$ $[\sin\theta(0)\cos\phi(0),\ \sin\theta(0)\sin\phi(0),\ \cos\theta(0)]$, $\boldsymbol{\sigma}$ = $(\sigma_x,\ \sigma_y,\ \sigma_z)$. General single-qubit gates can be constructed by selecting different parameters. Universal quantum computation can be realized by a non-trivial two-qubit quantum gate assisted with arbitrary single-qubit gates~\cite{PhysRevLett.89.247902}. 

For the two-qubit gate, the auxiliary bases can be chosen as 
\begin{eqnarray}
&&|\varphi_1(t)\rangle=|00\rangle,\notag\\
&&|\varphi_2(t)\rangle=\cos{\frac{\theta(t)}{2}}|01\rangle+\sin{\frac{\theta(t)}{2}e^{i\phi(t)}}|10\rangle,\notag\\
&&|\varphi_3(t)\rangle=\sin{\frac{\theta(t)}{2}}e^{-i\phi(t)}|01\rangle-\cos{\frac{\theta(t)}{2}}|10\rangle,\notag\\
&&|\varphi_4(t)\rangle=|11\rangle.
\end{eqnarray}
The general form of Hamiltonian for realizing a two-qubit gate is 
\begin{eqnarray}
H_{\rm two}(t)&=&i\sum_{k\neq l}^{4}\langle \varphi_l(t)|\dot{\varphi}_k(t)\rangle|\varphi_l(t)\rangle\langle\varphi_k(t)|\notag\\
&=&\frac{\Omega_{\rm two}(t)}{2}|01\rangle\langle10|+\rm H.c..
\label{eq6}
\end{eqnarray}

The evolution operator of this system can be expressed as 
\begin{widetext}
	\begin{eqnarray}
U_2(T)&=&|\varphi_1(T)\rangle\langle\varphi_1(0)|+e^{-i\gamma(T)}|\varphi_2(0)\rangle\langle\varphi_2(0)|+e^{i\gamma(T)}|\varphi_3(0)\rangle\langle\varphi_3(0)|+\varphi_4(T)\rangle\langle\varphi_4(0)|\notag 
\\
&=&\begin{pmatrix}
    1&0&0&0\\    0&\cos{\gamma(T)}-i\cos{\theta(0)}\sin{\gamma(T)}&-ie^{-i\varphi(0)}\sin{\theta(0)}\sin{\gamma(T)}&0\\
    0&-ie^{i\varphi(0)}\sin{\theta(0)}\sin{\gamma(T)}&\cos{\gamma(T)}+i\cos{\theta(0)}\sin{\gamma(T)}&0\\
    0&0&0&1\\
 \end{pmatrix}.\notag\\
	\end{eqnarray}
\end{widetext}

 By selecting appropriate parameters, we can build a nontrivial two-qubit quantum geometric gate and the general geometric quantum computation can be achieved by combining it with universal single-qubit gates.

\bibliography{DFS}

\end{document}